\pdfoutput=1

\documentclass{llncs}
\usepackage[utf8]{inputenc}
\usepackage[nolist]{acronym}
\usepackage{amsfonts,epsfig,latexsym,amssymb,yhmath,hyperref}
\usepackage{graphicx}
\usepackage{booktabs}
\usepackage{array}
\usepackage{makeidx}  
\usepackage[usenames,dvipsnames]{color}
\usepackage{cleveref}
\usepackage{float}

\newcolumntype{L}[1]{>{\raggedright\let\newline\\\arraybackslash\hspace{0pt}}m{#1}}
\newcolumntype{C}[1]{>{\centering\let\newline\\\arraybackslash\hspace{0pt}}m{#1}}
\newcolumntype{R}[1]{>{\raggedleft\let\newline\\\arraybackslash\hspace{0pt}}m{#1}}

\usepackage{makeidx}  
\usepackage[usenames,dvipsnames]{color}
\DefineNamedColor{named}{cornflower}{cmyk}{1,0.7,0,0}
\DefineNamedColor{named}{Lila}{cmyk}{0.69,0.83,0,0}
\DefineNamedColor{named}{Orange}{cmyk}{0.01,0.5,0.97,0}
\DefineNamedColor{named}{Red}{cmyk}{0,0.999,0.99,0}
\DefineNamedColor{named}{RedOrange}{cmyk}{0,0.77,0.87,0}
\DefineNamedColor{named}{LimeGreen}{cmyk}{0.50,0,1,0}
\DefineNamedColor{named}{Green}{cmyk}{1,0,1,0}
\DefineNamedColor{named}{ForestGreen} {cmyk}{0.91,0,0.88,0.12}
\DefineNamedColor{named}{Emerald}{cmyk}{1,0,0,0}
\DefineNamedColor{named}{Blue}{rgb}{0,0,1}
\DefineNamedColor{named}{DarkEmerald}{cmyk}{1,0,0,0.4}
\DefineNamedColor{named}{Maroon}{cmyk}{0,0.87,0.68,0.32}
\DefineNamedColor{named}{Fuchsia} {cmyk}{0.47,0.91,0,0.08}
\DefineNamedColor{named}{RealRed}{cmyk}{0,1,0.98,0}
\DefineNamedColor{named}{Gray}{cmyk}{0,0,0,0.6}

\newif\ifshowrevisionchanges
\showrevisionchangesfalse

\title{Data-centric AI approach to improve optic nerve head segmentation and localization in OCT en face images}

\author{
Thomas Schlegl\inst{1}\thanks{Corresponding author: Thomas Schlegl, Center for Medical Physics and Biomedical Engineering, Medical University of Vienna, Waehringer Guertel 18-20, 1090, Vienna, Austria. thomas.schlegl@meduniwien.ac.at} \and Heiko Stino\inst{2} \and Michael Niederleithner\inst{1} \and Andreas Pollreisz\inst{2} \and Ursula Schmidt-Erfurth\inst{2} \and Wolfgang Drexler\inst{1} \and Rainer A. Leitgeb\inst{1} \and Tilman Schmoll\inst{1,3}
}
\institute{$^1$Center for Medical Physics and Biomedical Engineering, Medical University of Vienna, Vienna, Austria. $^2$Department of Ophthalmology, Medical University of Vienna, Vienna, Austria. $^3$Carl Zeiss Meditec, Inc., Dublin, CA, USA}

\begin{document}

\maketitle

\begin{abstract}
The automatic detection and localization of anatomical features in retinal imaging data are relevant for many aspects. In this work, we follow a data-centric approach to optimize classifier training for optic nerve head detection and localization in optical coherence tomography en face images of the retina.
We examine the effect of domain knowledge driven spatial complexity reduction on the resulting optic nerve head segmentation and localization performance. 
We present a  machine learning approach for segmenting optic nerve head in 2D en face projections of 3D widefield swept source optical coherence tomography scans that enables the  automated assessment of large amounts of data.
Evaluation on manually annotated 2D en face images of the retina demonstrates that training of a standard U-Net can yield improved optic nerve head segmentation and localization performance when the underlying pixel-level binary classification task is spatially relaxed through domain knowledge.

\end{abstract}

\hypersetup{pdfauthor={Thomas Schlegl},pdftitle={Data-centric AI approach to improve optic nerve head segmentation and localization in OCT en face images}}

\section{Introduction}
In this paper, we propose to leverage a deep neural network for automated segmentation of the ONH in 2D en face projections of 3D widefield (60 degrees) swept source OCT (SS-OCT) scans. We utilize a simple U-Net architecture for accurate ONH detection and localization, described in Section~\ref{sec:methods:sem_seg}.\\
The acquisition of retinal OCT scans is standardized. Therefore, the rough location of the ONH and of other anatomical structures (like the fovea) is known. The overall image content is only flipped on the vertical axis depending whether the en face image has been reconstructed from a 3D OCT scan of the left or of the right eye.
Therefore, the image regions where the ONH can be located in the en face images is a priori known to a large extent.
A spatially unrestricted image has a larger field of view and therefore comprises a larger amount of image content. Images with a large field of view, in turn, need a large receptive field of the machine learning model. Both challenges require an increased machine learning model capacity and call for optimized training strategies that are typically needed to empower the machine learning model to learn a large variability of anatomical and pathological medical images. 
In this work, we aim to improve classifier training without the need to simultaneously increase the computational burden in terms of model complexity, run time, memory footprint, or the need for extensive data augmentation tweaks. We propose to simply spatially restrict the data fed to the classifier during training and during inference.
In the spirit of data-centric artificial intelligence (AI), we propose to apply domain specific data modifications instead of extensively tune machine learning architectures and related learning strategies.
Data-centric AI is the recently evolved discipline of focusing the engineering work and research specifically on the data that is used for training a machine learning model at least as much as on tuning of the machine learning model itself. The data-centric AI discipline shifts the data engineering task from the past one-time preprocessing approach to a systematically engineering of the data realized by an iterative \textit{data tuning} process that is an integral component of the training-validation cycle.

\paragraph{Clinical background and motivation}
The optic nerve head (ONH) is an important anatomical structure in several aspects. It constitutes a distinct anatomical structure located in the posterior segment of the eye and comprises blood vessels, ganglion cell axons, connective tissue and glia~\cite{reis2012optic}. In color fundus images as well as in OCT en face images, the area of the optic nerve head is typically depicted with distinct different brightness or color values compared to the color or gray values of the surrounding anatomical structures.

Often the detection and localization of the ONH in 2D en face view serves as a preprocessing step for subsequent image analysis steps even if the subsequent analysis is performed in the original 3D OCT volume space.
Automated localization of the ONH enables the implementation of a fully automated image processing pipeline where an ONH mask is required as an integral component.
Here, we apply supervised machine learning and deep neural networks for accurate fully automated ONH localization.

\paragraph{Related Work}
In contrast to applying convolutional neural networks (CNN) on image patches to perform segmentation of a full image through densely extracting and classifying those image parts,  deep learning based semantic segmentation approaches, such as \textit{fully convolutional networks (FCN)} proposed by Long et al.~\cite{long2015fully} or \textit{DeepLab} proposed by Chen et al.~\cite{chen2015semantic}, yield higher segmentation accuracy with a reduced computational cost.
Ronneberger et al.~\cite{ronneberger2015u} introduced the U-Net model architecture and won the \textit{International Symposium on Biomedical Imaging (ISBI)} cell tracking challenge (2015) by training this network on transmitted light microscopy images. Today the U-Net is one of the state-of-the art deep learning architectures for a wide range of image segmentation tasks.  
The U-Net architecture enables fast end-to-end training and inference from input images to dense label maps of the same resolution. Furthermore, it can be trained from only a few training samples~\cite{ronneberger2015u}.

The most closely related work to our approach was presented by Fard and Bagherinia~\cite{fard2019automatic}, who developed an automatic optic nerve head detection algorithm in widefield OCT using an U-Net and subsequent template matching strategy, in order to find the ONH center using a $4 mm$ diameter disc-shaped binary mask based on the U-Net output. As 3-channel input to the U-Net, an OCT en face image, a vessel enhanced OCT en face, and an OCT contrast map were used. The approach is evaluated based on the \textit{ONH center} localization accuracy.
Our work differs in the following aspects: although the euclidean distance between the ground-truth and the predicted ONH center is used in our work as performance metric, \textbf{(i)} we do not principally aim to predict the ONH center but we primarily tackle an ONH segment task, \textbf{(ii)} instead of relying on 3-channel inputs only a plain 2D OCT en face image is used as single network input, and \textbf{(iii)} no template matching strategy has to be utilized on the output of the neural network prediction as a postprocessing step. Both, the reduced computational cost for the computation of the inputs and the end-to-end learning and direct mapping from an input image to the corresponding prediction of the pixel-level ONH region during inference render our approach overall simpler and faster to train and to apply on new data.

\paragraph{Contribution}
We propose to adopt a data-centric AI strategy, namely to simply restrict the spatial complexity (i.e., the size of the input images) to the relevant image regions instead of exceedingly tune the utilized model architecture, model complexity, data augmentation strategies, or the therewith related training strategies. 
Since the patient positioning and image acquisition of the underlying 3D OCT scan is standardized and the variability of the ONH location relative to other anatomical structures in the retina is limited, we can perform this spatial complexity reduction through simply cropping the input images.
By cropping the input images to the region of the most probable location of the ONH, the variability of the image content is reduced as well.\\
We evaluate the effect of different input image sizes on the resulting segmentation accuracy and study whether we see general improvement of the achievable model performance irrespective of the specific training strategy by using different training objectives.

Experiments (Section~\ref{sec:experiments}) on annotated 2D en face projections of 3D OCT scans show that the proposed approach segments and localizes the ONH in 2D en face images with high accuracy.
To the best of our knowledge, this is the first published work on fully automated detection and localization of the ONH in 2D en face projections of 3D widefield SS-OCT scans that tunes the model performance primarily via data engineering and that additionally only utilizes a single neural network without the need for refinement in a postprocessing step.

\section{Data-centric segmentation of the optic nerve head}
In this work we focus on the benefits and effects of data-centric engineering. We utilize a machine learning segmentation model to perform and evaluate the proposed approach.

We apply semantic segmentation~\cite{mostajabi2015feedforward,noh2015learning,long2015fully,chen2015semantic,zheng2015conditional} to perform pixel-level classification in a single pass in order to detect and localize the object of interest in an image. As opposed to performing image segmentation based on an image-level classification on small image patches, semantic segmentation is very run-time efficient. 

In the following, we describe the data, the utilized semantic segmentation approach, and the corresponding training objectives in more detail.

\subsection{Data representation}
\label{sec:methods:representations}
The data comprises $N$ tuples of medical images and pixel-level ground truth annotations
$\langle \mathbf{X}_n, \mathbf{Y}_n \rangle$, with $n=1,2,\dots ,N$, where $\mathbf{X}_n \in \mathbb{R}^{a \times b}$ is an intensity image of size $a \times b$ and
$\mathbf{Y}_n \in \{0,1\}^{a \times b}$ is a binary image of the same size specifying the pixel-level presence of the object of interest.\\
The data is divided into disjoint sets, used for training, validation, or testing of the model.

\subsection{Data-centric optimization}
\label{sec:methods:dataCentric}
For data-centric performance optimization of the utilized machine learning model, we reduce the spatial size of both, the intensity input image $\mathbf{X}_n \in \mathbb{R}^{a \times b}$ and of the binary target image $\mathbf{Y}_n \in \{0,1\}^{a \times b}$.\\
We crop the 2D intensity input images $\mathbf{X}_n$ and the corresponding binary target images $\mathbf{Y}_n$ to the size $\dot{a} \times \dot{b}$, with $\dot{a} \leq a$ and $\dot{b} \leq b$ so that for training of the machine learning model we use tuples of intensity input images $\mathbf{X}_n \in \mathbb{R}^{\dot{a} \times \dot{b}}$ and corresponding binary target images $\mathbf{Y}_n \in \{0,1\}^{\dot{a} \times \dot{b}}$.
During inference, inputs of the same (reduced) image size $\dot{a} \times \dot{b}$ have to be fed to the machine learning model.

\subsection{Semantic segmentation methodology}
\label{sec:methods:sem_seg}

\paragraph{U-Net based semantic segmentation}
We leverage an \textbf{\textit{U-Net}} architecture to learn the mapping $M(\mathbf{X}) = \mathbf{X} \mapsto \mathbf{Y}$ from intensity images $\mathbf{X}$ to corresponding binary images $\mathbf{Y}$ of dense pixel-level class labels by training a deep neural network $M$. During testing, the model $M$ yields images $\mathbf{\hat{Y}}$ of dense pixel-level predictions for unseen testing images $\mathbf{X}_u$.
The U-Net architecture comprises a contracting path (\textit{encoder}) and an expanding path (\textit{decoder}), which are jointly trained. The encoder transforms the input image into a low-dimensional abstract representation of the image content. This low-dimensional embedding is fed into the \textit{decoder}, which maps this representation to corresponding dense class label predictions of a full input resolution. Convolutional layers are the main processing units of encoder and decoder. 
The utilization of encoder features in the decoder, referred to as \textit{skip connections}, has been one of the main contributions of the U-Net architecture proposed by Ronneberger et al.~\cite{ronneberger2015u}.
Based on tuples of intensity images and corresponding dense target labels, the semantic segmentation model parameters of the encoder and of the decoder are updated in every update iteration (\textit{end-to-end training}).

\subsection{Training objectives}
\label{sec:methods:ojectives}
We train networks separately on two objective functions, on \textit{binary cross entropy loss} or on \textit{Tversky loss}. At each pixel location, both loss functions penalize the deviation of the network prediction from the ground truth binary class labels.

\subsubsection{Binary cross entropy loss}
The (pixel-level) binary cross entropy loss is computed between class probabilities $\hat{Y}_i$ and target labels $Y_i$:
\begin{equation}\label{eqn:binary_ce_loss}
	L_i =  -Y_i \log(\hat{Y}_i) - (1-Y_i) \log(1-\hat{Y}_i),
\end{equation}
where $\hat{Y}_i$ is the sigmoidal output $M(X_i)$ of the neural network $M$ for the i-th pixel in image $X$
\begin{equation}\label{eqn:sigmoid}
	\sigma(z_i) = \frac{1}{1 + e^{-z_i}}
\end{equation}

\subsubsection{Tversky loss}
For binary segmentation problems, the \textit{Dice similarity coefficient (DSC)}~\cite{sorensen1948method,dice1945measures} quantifies the ``similarity'' of predicted and the true segmentation, and is defined by:
\begin{equation}\label{eqn:dice_score}
	DSC = \frac{2 \cdot t^+}{2 \cdot t^+ + f^+ + f^-},
\end{equation}
where $t^+$ is the number of \textit{true positives}, $f^+$ is the number of \textit{false positives}, and $f^-$ is the number of \textit{false negatives}. Values of DSC range from 0.0 (worst classification) to 1.0 (perfect classification).

The Tversky index~\cite{tversky1977features} $T$ can be interpreted as a generalization of the Dice similarity coefficient by introducing parameter $\beta$:
\begin{equation}\label{eqn:tversky_index}
	T = \frac{t^+}{t^+ + \beta \cdot f^+ + (1-\beta) \cdot f^-}.
\end{equation}

For binary segmentation problems, the Tversky index can not only be utilized to evaluate the performance of a trained model on the test set, but also as basis for an objective function during training. We can define a \textit{Tversky loss} $T_L$ objective function to train a model to minimize:
\begin{equation}\label{eqn:ce_loss}
	T_L = 1.0 - \frac{t^+ + \epsilon}{t^+ + \beta \cdot f^+ + (1-\beta) \cdot f^- + \epsilon}
\end{equation}
between real-valued network predictions $\mathbf{\hat{Y_i}}$ and binary target labels $\mathbf{Y_i}$ for the i-th pixel, where $\epsilon$ is a small real-valued constant to prevent numerical instabilities.

\section{Experiments}
\label{sec:experiments}
ONH appear in OCT en face images as brighter or darker disks compared to the surrounding structures. ONH segmentation can be formulated as pixel-wise binary classification problem. We examine whether the reduction of the segmentation task complexity through domain knowledge while holding constant the utilized model complexity can improve the segmentation and detection performance. In concrete terms, we apply the same model on the full en face images and on two gradually cropped versions of the image data.

\paragraph{Data, Data Selection and Preprocessing}
For model training and evaluation we used a dataset comprising a total of 120 annotated OCT en face images with an image resolution of $2048 \times 2048$ pixels (pixel sizes $8.8 \times 8.8\mu m$), from 89 eyes of 64 diabetic retinopathy patients. 
We split the data into a training set of 100 en face images (70 unique eyes from 48 patients), a validation set of 10 en face images (9 unique eyes from 7 patients),
and a test set of 10 en face images (10 unique eyes from 9 patients).  For 30 patients of the training set and for 1 patient of the validation set, the dataset comprises more than one scan per eye. Mainly caused by patient related limited scanning conditions, some of the scans have reduced image quality.
We performed model training on the training set. The validation set was used for hyper-parameter tuning and model selection. The test set was only used once, namely, for the final model evaluation. We split the data on the patient-level, so that scans of individual patients are exclusively contained in the training set, validation set, or test set.
The en face images represent mean gray values along the z-axis of 3D OCT scans acquired with an OCT prototype that has been developed by our group~\cite{niederleithner2020clinical}.
The used system is a swept-source OCT (SS-OCT) system utilizing a Fourier-domain mode-locking (FDML) laser with a sweep-rate (A-scan rate) of $1.68 MHz$, a central wavelength of $1060 nm$ and $75 nm$ tuning range, translating to an axial resolution of $9\mu m$ in tissue. The used field of view is $18 mm$ in diameter sampled with $2048 \times 2048$ A-scans per volume, resulting in a sampling density of around $8.8 \mu m$. With a beam diameter of $1 mm$ on the pupil, the lateral resolution on the retina is approximately $20\mu m \: (1/e^2)$, therefore fulfilling the Nyquist's criterion.

Among patients, there is low variability in relative size of the ONH compared to the full en face image. Additionally, when downscaling the original en face images by a factor of 8, the ONH can still be identified. Therefore, we perform ONH segmentation and detection on downscaled en face images with image resolution of $256 \times 256$ pixels. Other than that, only the input gray values were normalized to range from 0 to 1.

\paragraph{Image size reduction for optimized machine learning performance}
Based on the inferences we drew from data exploration and visualization on the samples of the training set about the spatial distribution of the optic nerve head, we only reduce the number of columns of the images. In our experiments we used three different sizes of 2D intensity input images and corresponding binary target images, namely images of size $160 \times 256$ pixels (\textit{moderate spatial reduction}), $96 \times 256$ pixels (\textit{significant spatial reduction}), and -- as baseline -- images with $256 \times 256$ pixels (\textit{no cropping}).

\subsection{Evaluation}
\label{sec:exp_eval}

\paragraph{Semantic segmentation model}
We evaluate the semantic segmentation performance of a standard \textit{U-Net} model with convolutional layers as main units, where the encoder and decoder comprise six and five blocks of two convolutional layers with $(16-32-64-128-256-512)$ and $(256-128-64-32-32)$ filters of size $(3 \times 3)$ pixels, respectively. The first five encoder blocks are followed by a max pooling layer. Whereas, dropout is applied on the output of the sixth encoder block with a dropout rate of $0.2$.
While keeping the model complexity constant, we perform model training with two different loss functions, namely \textbf{(1)} \textit{binary cross entropy loss} or \textbf{(2)} \textit{Tversky loss}, to examine whether we observe general influence of spatial image complexity on the segmentation performance for both loss functions.

\paragraph{Evaluation metrics}
We use the Dice similarity coefficient (\textit{Dice}) to evaluate the pixel-level segmentation performance as a measure for the \textbf{ONH detection accuracy} of a model.
In skewed class distributions of the data, receiver operating characteristic (ROC) curves can present a too optimistic visualization of the model's \textit{binary classification} performance~\cite{davis2006relationship}. For the evaluation of the classification performance on data with skewed class distribution, precision-recall curves are an appropriate alternative to ROC curves ~\cite{goadrich2004learning,bunescu2005comparative,craven2005markov}.
Binary pixel values of a segmentation mask that localize the optic disc in the corresponding gray value en face image represent highly imbalanced classes, i.e. the coverage of the optic disc relative to the full total area of the en face image is relatively small.
Therefore, we use \textit{precision-recall curves} to visualize the model performance on the ONH segmentation task. The \textit{average precision (aPr)} of the model
is a quantitative summary statistic for the precision-recall curve.
The Dice similarity coefficient and further quantitative performance statistics (sensitivity, specificity, and precision) are calculated at the optimal cut-off point of the precision-recall curve. We report the area under the ROC curve (AUC) values for the sake of completeness only.\\

Furthermore, we evaluate the ONH \textbf{localization performance} based on the Euclidean distance between the centroids of the pixel-level ground truth annotation mask and the corresponding predicted ONH segmentation result.

\paragraph{Implementation details}
All models were trained for 300 epochs, and model parameters were stored at the best performing epoch on the validation set. After model selection and hyperparameter tuning, the final performance was evaluated on the test set using the learned model parameters. We utilized the stochastic optimizer Adam~\cite{kingma2014adam} during training. All experiments were performed using Python 3.8 with the TensorFlow~\cite{tensorflow2015-whitepaper} library version 2.2 and the high-level API Keras~\cite{chollet2015keras} 2.7, CUDA 11.4, and a NVIDIA Titan Xp graphics processing unit.

\subsection{Results}
\begin{figure}[htp]
  \centering
     \includegraphics[width=1\textwidth]{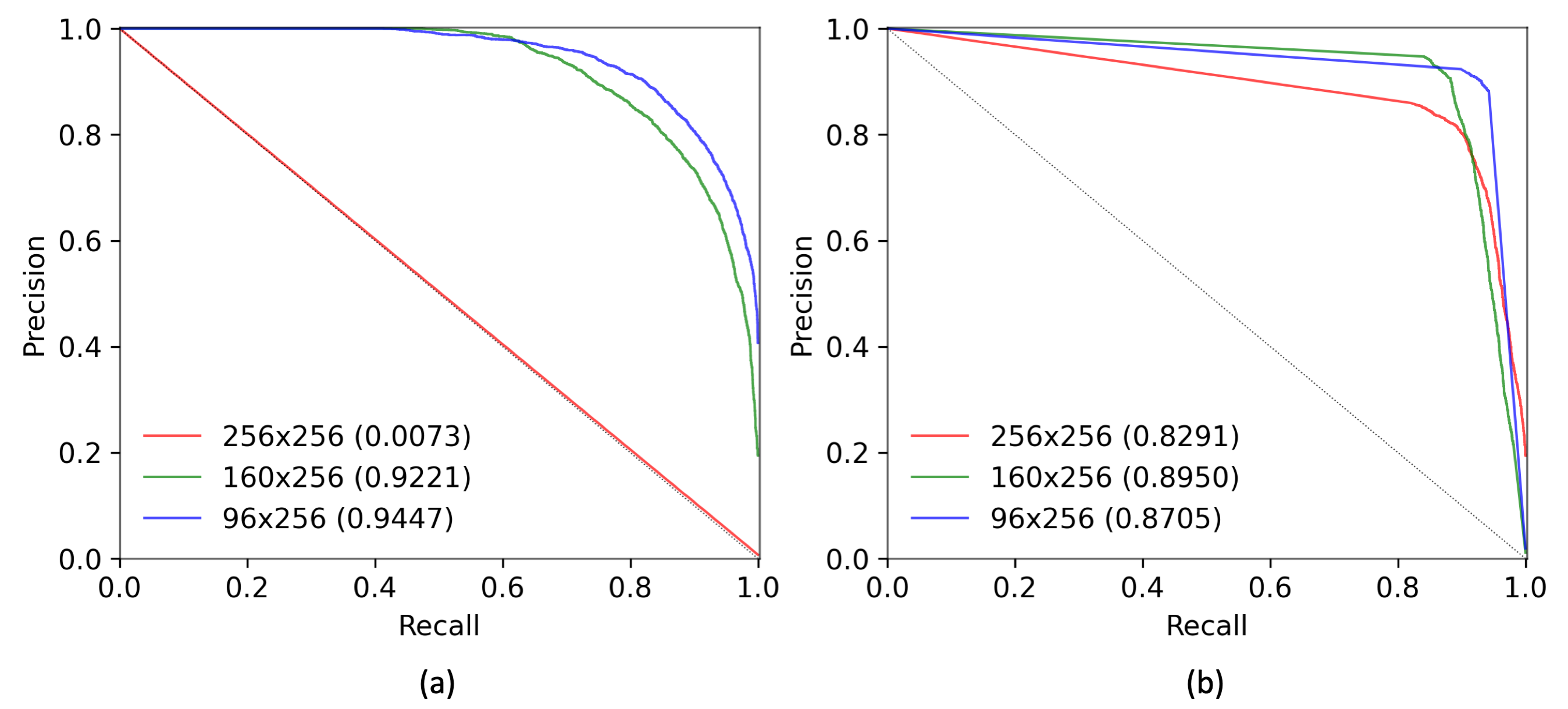}
     \vspace{-5mm}
  \caption{
  ONH segmentation performance evaluation. Precision-recall curves for models based on an U-Net with $256\times256$ pixel inputs, $160\times256$ pixel inputs, or $96\times256$ pixel inputs trained with different loss functions: a) binary cross entropy (bce) loss and b) Tversky loss.  Corresponding average precision (\textit{aPr}) values are given in parenthesis.
 }
  \label{fig:result/PRc}
\end{figure}

\begin{figure}[htp]
  \centering
     \includegraphics[width=1\textwidth]{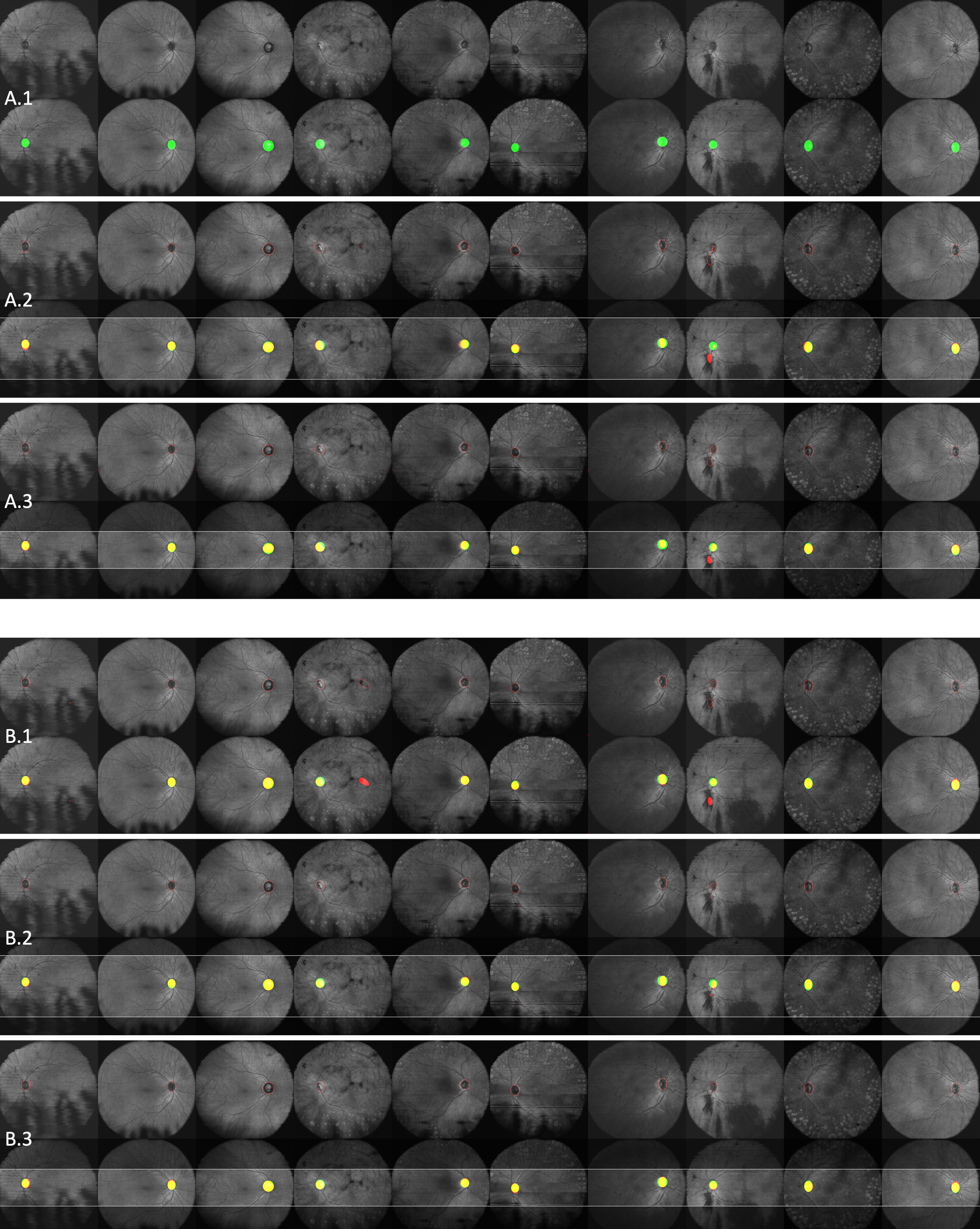}
  \caption{ONH segmentation results on test set samples.  Pixel level segmentation results based on an U-Net with $256\times256$ pixel inputs (A.1 and B.1), $160\times256$ pixel inputs (A.2 and B.2), or $96\times256$ pixel inputs (A.3 and B.3). (A.1-3) Models trained utilizing a binary cross entropy (bce) loss. (B.1-3) Models trained utilizing a Tversky loss.
En face images with resulting ONH segmentation boundaries (first rows) and en face images overlayed with corresponding pixel-level segmentation results (second rows). True positives (yellow), false positives (red), false negatives (green), and true negatives (gray values of en face image). The input images are cropped to the central image region delineated by white horizontal lines overlayed on the en face images in the bottom rows.}
\label{fig:result/seg_result_imgs}
\end{figure}

Results demonstrate that the spatial restriction of the input images improves resulting model performance for both tasks, ONH segmentation and ONH localization. In our experiments, this beneficial effect is independent of the specific loss function used for training.
Detailed quantitative results are listed in Table~\ref{tab:result/quant_results}, which show that the model trained with the Tversky loss on $96 \times 256$ input images yields best mean Dice score of $0.9154$ evaluated on the test set, which is a measure for the overall segmentation and detection accuracy of the approach. Furthermore, it performs best with regard to the Euclidean distance (eDist of $1.2730$ pixels) between the centroids of the pixel-level ground truth annotation mask and the corresponding predicted ONH segmentation result, which is a measure of the ONH localization accuracy.
The observation that a restriction of the input images to the relevant image part improves resulting segmentation and localization performance also holds true for the models trained with BCE loss, with increased Dice score of $0.8615$ and decreased Euclidean distance of $2.6995$ compared to a model trained on $160 \times 256$ input images, where the model yields a Dice score of $0.8305$ and a Euclidean distance of $3.8575$. On the full-sized input images a model trained with BCE loss even completely fails to detect the ONH.
The benefit of model training on spatially restricted input images is also evident through the precision-recall curves shown in Figure~\ref{fig:result/PRc}. Qualitative segmentation results are separately shown for models trained with BCE~\Cref{fig:result/seg_result_imgs} (A.1-3) and with Tversky loss~\Cref{fig:result/seg_result_imgs} (B.1-3) for the three input sizes.\\
On the GPU, the mean inference times for all input image resolutions were lying in the magnitude of $0.007 sec$.

\begin{table}[H]
\centering
\caption{
Evaluation results on the test set for models trained utilizing a binary cross entropy loss (\textit{bce}) or utilizing a Tversky loss (\textit{tversky}) - trained and evaluated on $256\times256$ pixel inputs, $160\times256$ pixel inputs, or $96\times256$ pixel inputs with 256, 160, or 96 input rows (\textit{ir}), respectively.
The quantitative performance statistics (sensitivity, specificity, and precision) and Dice similarity coefficient (\textit{Dice}) measuring the pixel-level segmentation performance calculated at the optimal cut-off point of the precision-recall curve,  the corresponding average precision (\textit{aPr}), and the mean Euclidean distance (\textit{eDist}) between the centroids of the pixel-level ground truth annotation mask and the corresponding predicted ONH segmentation result measuring the ONH localization accuracy. Due to the strong skewness of the pixel-level class label distribution, we report the area under the receiver operating characteristic (\textit{ROC}) curve (\textit{AUC}) only for completeness.
}
\vspace{-2mm}
    \begin{tabular}
    { @{}L{.07\textwidth} L{.09\textwidth} | C{0.13\textwidth}  C{0.12\textwidth}  C{0.11\textwidth}  C{0.09\textwidth} C{0.09\textwidth} | C{0.09\textwidth} | C{0.09\textwidth} }
ir  &  loss  &  sensitivity  &  specificity  &  precision  &  AUC  &  aPr  &   Dice  &  eDist \\ \hline
256  &  bce  &  0.0000  &  1.0000  &  0.0000  &  0.5000  &  0.0073  &  0.0000  &  nan  \\ 
160  &  bce  &  0.8280  &  0.9980  &  0.8331  &  0.9986  &  0.9221  &  0.8305  &  3.8575  \\ 
 96  &  bce  &  0.8382  &  0.9979  &  0.8861  &  0.9987  &  0.9447  &  \textit{0.8615}  &  \textit{2.6995}  \\ \hline  
256  &  tversky  &  0.8921  &  0.9985  &  0.8155  &  0.9988  &  0.8291  &  0.8521  &  8.7934  \\ 
160  &  tversky  &  0.8516  &  0.9993  &  0.9392  &  0.9889  &  0.8950  &  0.8933  &  1.5614  \\ 
 96  &  tversky  &  0.9303  &  0.9980  &  0.9010  &  0.9704  &  0.8705  &  \textbf{0.9154}  &  \textbf{1.2730}  \\ 
    \end{tabular}
\normalsize
\label{tab:result/quant_results}
\vspace{-6mm}
\end{table}

\section{Conclusion} 
We applied neural network based binary segmentation for fully automated detection and localization of ONH in OCT en face images. We showed that a spatial restriction of the input images to a specific area where the object of interest (in our case the ONH) can occur, improves the resulting model performance on the ONH detection and localization task.\\
However, such restriction always requires domain knowledge for the specific task at hand. In many cases where the object of interest is not located at the exactly same location of the image a priori, a spatial restriction to a specific area of the inputs can be achieved via spatial normalization of the images.
The question to what extent the beneficial effect of spatial normalization of the inputs is comparable to an increased model complexity and -- as a consequence thereof -- a necessarily increased number of training samples is left for future research.\\
Some experiments -- especially the scenarios with spatially unrestricted input images -- lead to oversegmentations, which have a strong influence on the Euclidean distance metric as a measure for the ONH localization accuracy. This problem can be handled in a postprocessing step by only keeping the largest connected component. The study of neural network architectures that are capable of eliminating oversegmentations from the beginning is left to future work.\\
For all experiments we used the same simple neural network architecture. When performing training and evaluation on spatially unrestricted images, the chosen network architecture doesn't optimally generalize from training images to validation images. When utilizing a Tversky loss the model performance on spatially unrestricted images of the test set is inferior to a model of the same complexity trained on spatially restricted input images. In the case of model training with a BCE loss, the model entirely fails to segment the ONH but performs comparably well on spatially restricted input images.
A restriction of the inputs not only reduces the extend of false positive predictions but also improves the general segmentation accuracy, which is reflected by an increased Dice score.\\
Those consistent results on two different loss functions suggest that a spatial restriction of the inputs can foster neural network training in general.

\section*{Acknowledgements}
This work has received funding from the European Commission H2020 program initiated by the Photonics Public Private Partnership (MOON number 732969).

\bibliographystyle{splncs}
\bibliography{references}

\begin{thebibliography}{10}

\bibitem{reis2012optic}
Reis, A.S., Sharpe, G.P., Yang, H., Nicolela, M.T., Burgoyne, C.F., Chauhan,
  B.C.:
\newblock Optic disc margin anatomy in patients with glaucoma and normal
  controls with spectral domain optical coherence tomography.
\newblock Ophthalmology \textbf{119}(4) (2012)  738--747

\bibitem{long2015fully}
Long, J., Shelhamer, E., Darrell, T.:
\newblock Fully convolutional networks for semantic segmentation.
\newblock In: Proceedings of the IEEE conference on computer vision and pattern
  recognition. (2015)  3431--3440

\bibitem{chen2015semantic}
Chen, L.C., Papandreou, G., Kokkinos, I., Murphy, K., Yuille, A.:
\newblock Semantic image segmentation with deep convolutional nets and fully
  connected {CRF}s.
\newblock In: International Conference on Learning Representations. (2015)

\bibitem{ronneberger2015u}
Ronneberger, O., Fischer, P., Brox, T.:
\newblock U-net: Convolutional networks for biomedical image segmentation.
\newblock In: International Conference on Medical image computing and
  computer-assisted intervention, Springer (2015)  234--241

\bibitem{fard2019automatic}
Fard, A., Bagherinia, H.:
\newblock Automatic detection of optic nerve head in widefield {OCT} using deep
  learning.
\newblock Investigative Ophthalmology \& Visual Science \textbf{60}(11) (2019)
  PB0100--PB0100

\bibitem{mostajabi2015feedforward}
Mostajabi, M., Yadollahpour, P., Shakhnarovich, G.:
\newblock Feedforward semantic segmentation with zoom-out features.
\newblock In: Proceedings of the IEEE conference on computer vision and pattern
  recognition. (2015)  3376--3385

\bibitem{noh2015learning}
Noh, H., Hong, S., Han, B.:
\newblock Learning deconvolution network for semantic segmentation.
\newblock In: Proceedings of the IEEE International Conference on Computer
  Vision. (2015)  1520--1528

\bibitem{zheng2015conditional}
Zheng, S., Jayasumana, S., Romera-Paredes, B., Vineet, V., Su, Z., Du, D.,
  Huang, C., Torr, P.H.:
\newblock Conditional random fields as recurrent neural networks.
\newblock In: Proceedings of the IEEE International Conference on Computer
  Vision. (2015)  1529--1537

\bibitem{sorensen1948method}
S{\o}rensen, T.:
\newblock A method of establishing groups of equal amplitude in plant sociology
  based on similarity of species and its application to analyses of the
  vegetation on danish commons.
\newblock Biol. Skr. \textbf{5}(4) (1948)  1--34

\bibitem{dice1945measures}
Dice, L.R.:
\newblock Measures of the amount of ecologic association between species.
\newblock Ecology \textbf{26}(3) (1945)  297--302

\bibitem{tversky1977features}
Tversky, A.:
\newblock Features of similarity.
\newblock Psychological review \textbf{84}(4) (1977)  327

\bibitem{niederleithner2020clinical}
Niederleithner, M., Britten, A., Ginner, L., Salas, M., Ren, H., Arain, M.A.,
  Williams, R.A., Drexler, W., Leitgeb, R.A., Schmoll, T.:
\newblock Clinical megahertz-{OCT} for ophthalmic applications (conference
  presentation).
\newblock In: Ophthalmic Technologies XXX. Volume 11218., SPIE (2020)  112180E

\bibitem{davis2006relationship}
Davis, J., Goadrich, M.:
\newblock The relationship between precision-recall and {ROC} curves.
\newblock In: Proceedings of the 23rd international conference on Machine
  learning, ACM (2006)  233--240

\bibitem{goadrich2004learning}
Goadrich, M., Oliphant, L., Shavlik, J.:
\newblock Learning ensembles of first-order clauses for recall-precision
  curves: A case study in biomedical information extraction.
\newblock In: International Conference on Inductive Logic Programming, Springer
  (2004)  98--115

\bibitem{bunescu2005comparative}
Bunescu, R., Ge, R., Kate, R.J., Marcotte, E.M., Mooney, R.J., Ramani, A.K.,
  Wong, Y.W.:
\newblock Comparative experiments on learning information extractors for
  proteins and their interactions.
\newblock Artificial intelligence in medicine \textbf{33}(2) (2005)  139--155

\bibitem{craven2005markov}
Craven, M., Bockhorst, J.:
\newblock Markov networks for detecting overalpping elements in sequence data.
\newblock In: Advances in Neural Information Processing Systems. (2005)
  193--200

\bibitem{kingma2014adam}
Kingma, D., Ba, J.:
\newblock Adam: A method for stochastic optimization.
\newblock arXiv:1412.6980 (2014)

\bibitem{tensorflow2015-whitepaper}
Abadi, M., Agarwal, A., Barham, P., Brevdo, E., Chen, Z., Citro, C., Corrado,
  G.S., Davis, A., Dean, J., Devin, M., Ghemawat, S., Goodfellow, I., Harp, A.,
  Irving, G., Isard, M., Jia, Y., Jozefowicz, R., Kaiser, L., Kudlur, M.,
  Levenberg, J., Man\'{e}, D., Monga, R., Moore, S., Murray, D., Olah, C.,
  Schuster, M., Shlens, J., Steiner, B., Sutskever, I., Talwar, K., Tucker, P.,
  Vanhoucke, V., Vasudevan, V., Vi\'{e}gas, F., Vinyals, O., Warden, P.,
  Wattenberg, M., Wicke, M., Yu, Y., Zheng, X.:
\newblock {TensorFlow}: Large-scale machine learning on heterogeneous systems
  (2015) Software available from tensorflow.org.

\bibitem{chollet2015keras}
Chollet, F.,  et~al.:
\newblock Keras.
\newblock \url{https://keras.io} (2015)

\end{thebibliography}
\end{document}